\documentclass[twoside,10pt]{article}

\usepackage{multicol}
\usepackage{amsmath}
\usepackage{amsfonts} 
\usepackage{cite}

\newcommand{\Ent}[1]{[\mkern - 2.5 mu [#1] \mkern - 2.5 mu ]}  

\begin{document}

%\draft

% \preprint{WUHEP-00-18}

\title{Numerical Evidence that the Perturbation Expansion \\ 
for a Non-Hermitian $\mathcal{PT}$-Symmetric Hamiltonian \\ 
is Stieltjes}

\author{Carl M. Bender \\
Department of Physics, Washington University \\ 
St. Louis MO 63130, USA \\
cmb@howdy.wustl.edu \\
\and
Ernst Joachim Weniger \\
Institut f\"ur Physikalische und Theoretische Chemie \\
Universit\"at Regensburg, D-93040 Regensburg, Germany \\
joachim.weniger@chemie.uni-regensburg.de}

\date{Submitted to Journal of Mathematical Physics \\
2 October 2000}

\maketitle

% \pacs{PACS numbers: 03.65-w,02.30.Lt,11.10.Jj}

\typeout{==> Abstract}
\begin{abstract}
\noindent 
Recently, several studies of non-Hermitian Hamiltonians having $\mathcal{PT}$
symmetry have been conducted. Most striking about these complex Hamiltonians is
how closely their properties resemble those of conventional Hermitian
Hamiltonians. This paper presents further evidence of the similarity of these
Hamiltonians to Hermitian Hamiltonians by examining the summation of the
divergent weak-coupling perturbation series for the ground-state energy of the
$\mathcal{PT}$-symmetric Hamiltonian $H=p^2+\frac{1}{4}x^2+i\lambda x^3$
recently studied by Bender and Dunne. For this purpose the first 193 (nonzero)
coefficients of the Rayleigh-Schr\"odinger perturbation series in powers of
$\lambda^2$ for the ground-state energy were calculated. Pad\'e-summation and
Pad\'e-prediction techniques recently described by Weniger are applied to this
perturbation series. The qualitative features of the results obtained in this
way are indistinguishable from those obtained in the case of the perturbation
series for the quartic anharmonic oscillator, which is known to be a Stieltjes
series.
\end{abstract}

\typeout{==> Section 1} 
\section{Introduction} 
\label{Sec:intro} 

Hamiltonians describing fundamental interactions traditionally possess
two symmetries, the continuous symmetry of the proper Lorentz group and
the discrete symmetry of Hermiticity. Lorentz invariance is a physical
requirement. Hermiticity is a useful mathematical constraint that
guarantees that the spectrum is real, although recent work shows that
Hermiticity is only a sufficient condition and is not necessary for the
reality of eigenvalues. From the assumptions of Lorentz invariance and
positivity of the spectrum of the Hamiltonian one can prove the ${\cal
PCT}$ theorem and thereby establish the physical symmetry of ${\cal
PCT}$ invariance.

What happens if we impose only the more physical symmetries of Lorentz
invariance and ${\cal PCT}$ invariance when we construct a Hamiltonian? 
The constraint of ${\cal PCT}$ invariance is weaker than Hermiticity, so
Hamiltonians having this property need not be Hermitian. While it has
not been proved, there is compelling analytical and numerical evidence
supporting the conjecture that, except when ${\cal PCT}$ symmetry is
spontaneously broken, the energy levels of such Hamiltonians are all
real and positive\cite{R1,R2}. The reality and positivity of the
spectrum is apparently a consequence of the ${\cal PCT}$ symmetry of $H$.

Many examples of ${\cal PCT}$-symmetric Hamiltonians in quantum field
theory have been studied\cite{R3,R4,R5,R6,R7,R8}. In quantum mechanics,
where the ${\cal C}$ operator is unity, many examples of ${\cal
PT}$-symmetric Hamiltonians have also been
studied\cite{R9,R10,R11,R12,S1,S2,S3,S4,S5,S6,S7}. A simple example of
such a quantum-mechanical Hamiltonian is $H=p^2+ix^3$. Hamiltonians such
as this may be regarded as {\it complex deformations} of conventional
Hermitian Hamiltonians. To understand this deformation we consider the
Hamiltonian
\begin{displaymath}
H \; = \; p^2 - (ix)^{2 + \epsilon} \, ,
\end{displaymath}
where $\epsilon\geq0$. When $\epsilon=0$, we have the conventional
harmonic oscillator Hamiltonian, whose spectrum is real and positive. As
$\epsilon$ increases from $0$, the entire spectrum of the Hamiltonian
smoothly deforms as a function of $\epsilon$ and remains real and
positive for all positive values of $\epsilon$. Thus, these theories are
in effect the analytic continuation of conventional quantum mechanics
into the complex plane.

These non-Hermitian theories exhibit some remarkable properties. Most
interesting is that the expectation value of the operator $x$ in quantum
mechanics (and of the field $\phi$ in quantum field theory) is {\it
nonzero} when $\epsilon>0$. This is true even for the $p^2-x^4$
Hamiltonian that one obtains at $\epsilon=2$ and it is also true for the
$-g\phi^4$ scalar quantum field theory. The $-g\phi^4$ quantum field
theory is particularly surprising because it has a positive real
spectrum and exhibits a nonzero value of $\langle \phi \rangle$. In
four-dimensional space-time it has a dimensionless coupling constant, is
renormalizable, and is asymptotically free (and thus nontrivial). It may
thus provide a useful setting to describe the Higgs particle\cite{R7}.

We are struck by the close similarity between the properties of
non-Hermitian $\mathcal{PT}$-symmetric quantum-mechanical Hamiltonians
and conventional Hermitian Hamiltonians. Moreover, in mathematical
terms, we are struck by the strong resemblance between self-adjoint
Sturm-Liouville problems and these new complex Sturm-Liouville
problems. The purpose of this paper is to present further evidence of
this strong similarity by investigating various aspects of Pad\'{e}
summation and Pad\'{e} prediction of the Rayleigh-Schr\"odinger
perturbation series for the ground-state energy of the complex
$\mathcal{PT}$-symmetric Hamiltonian
\begin{equation}
H (\lambda) \; = \; p^2 + \frac{1}{4} x^2 + i \lambda x^3 \, . 
\label{e1} 
\end{equation} 
Note that this Hamiltonian is $\mathcal{PT}$ symmetric because under
parity reflection ${\cal P}:~p\to-p$ and ${\cal P}:~x\to-x$ and under
time reversal, which is an antiunitary operation, ${\cal T}:~p\to-p$,
${\cal T}:~x\to x$, and ${\cal T}:~i\to-i$.

The large-order behavior of the divergent Rayleigh-Schr\"odinger
perturbation series 
\begin{equation}
E_0 (\lambda) \; \sim \; \frac{1}{2} \, + \, 
\sum_{n=1}^\infty \, b_n \, \lambda^{2n} \, \qquad (\lambda \to 0^+)\, , 
\label{e2}
\end{equation} 
for the ground-state energy eigenvalue of the Hamiltonian in (\ref{e1})
has already been examined in \cite{R13}, where the first 46 terms of the
perturbation expansion had been generated using recursion formulas. It
was observed there that the coefficients $b_n$ are all integers, that
they alternate in sign, and that their magnitude grows rapidly with
$n$. The first 10 coefficients are listed in Table \ref{Table_1}.

To calculate the coefficients $b_n$ we make the {\em ansatz} that the wave
function is a formal series in powers of the coupling constant $\lambda$ and
that the coefficient of $\lambda^n$ has the form of a Gaussian $\exp(-x^2/4)$
times a polynomial of degree $3n$ in the variable $x$. The eigenvalue $E_0(
\lambda)$ automatically appears as a series in powers of $\lambda^2$. Thus, for
each additional coefficient in the series for $E_0$ it is necessary to calculate
{\em two} orders in powers of $\lambda$ for the wave function.

\begin{table}[ht] 
\begin{center}
\parbox{4.78in}{\caption{\label{Table_1} The first 10 
coefficients $b_n$ in the perturbation 
expansion (\protect\ref{e2}) for the ground state energy of the complex 
$\mathcal{PT}$-symmetric Hamiltonian (\protect\ref{e1}).}}
\begin{tabular*}{4.78in}{l@{\extracolsep{2.65in}}l} \\
[1\jot] \hline \hline \rule{0pt}{4\jot}%
$n$  & ${\phantom{-}}b_n$ \\ [1\jot] \hline \rule{0pt}{4\jot}%
$1 $ & ${\phantom{-}}11$ \\
$2 $ & $-930$ \\
$3 $ & ${\phantom{-}}158836$ \\
$4 $ & $-38501610$ \\
$5 $ & ${\phantom{-}}11777967516$ \\
$6 $ & $-4300048271460$ \\
$7 $ & ${\phantom{-}}1815215203378344$ \\
$8 $ & $-868277986898581530$ \\
$9 $ & ${\phantom{-}}464025598165231889260$ \\
$10$ & $-274145574452876905074540$ \\ 
[1\jot] \hline \hline \rule{0pt}{4\jot}%
\end{tabular*}
\end{center}
\end{table}

In \cite{R13} it was pointed out that the Hamiltonian (\ref{e1})
describes a $0+1$ dimensional $\phi^3$ field theory and that $\phi^3$
theories were the first quantum field theories in which the divergences
of perturbation theory were studied \cite{R14}. Using the standard
methods for determining the large-order behavior of perturbation theory
\cite{R15,R16} it can be shown that the leading large-$n$ behavior of
the coefficients $b_n$ is given by the formula
\begin{equation}
b_n \; = \; (-1)^{n+1} \, 60^{n+1/2}(2\pi)^{-3/2} \,\Gamma (n+1/2) \, 
\bigl[1 + {\rm O} \left( 1/n \right) \bigr] \, \qquad  (n \to \infty) \, .
\label{e3}
\end{equation}
This asymptotic behavior was verified numerically in \cite{R13}. There,
it was also shown that the first correction term to this leading
asymptotic behavior, which is proportional to $1/n$, is negative. 
Although divergent, the series in (\ref{e2}) is Borel summable
\cite{Bor1928,R17,R18,ShaWat1994}. If the factor of $i$ were absent from
the Hamiltonian (\ref{e1}), then the perturbation coefficients $b_n$
would not alternate in sign and the perturbation series would not be
Borel summable.

It is interesting that to derive the asymptotic formula in (\ref{e3}) one must
use dispersion-relation techniques that rely on crucial assumptions about the
analyticity of the function $E_0(\lambda)$. These assumptions are justified for
the Hermitian Hamiltonian of the anharmonic oscillator \cite{Si1970,HeSi1978},
\begin{equation}
\mathcal{H} (\beta) \; = \; p^2 \, + \, x^2 \, + \, \beta x^4 \, .
\label{DefQAHO}
\end{equation}
However, the validity of these assumptions is unproved for the
non-$\mathcal{PT}$-symmetric $\lambda x^3$ oscillator.

The eigenvalues $\mathcal{E}(\beta)$ of the quartic anharmonic oscillator
possess several other properties which closely resemble those of the eigenvalues
$E (\lambda)$ of the $\mathcal{PT}$-symmetric Hamiltonian (\ref{e1}). For
example, the ground-state energy eigenvalue $\mathcal{E}_0(\beta)$ of the
quartic anharmonic oscillator possesses a divergent weak-coupling perturbation
expansion, which also diverges factorially \cite{R15,WeCiVi1993}:
\begin{eqnarray}
\mathcal{E}_0 (\beta) & \sim & 1 \, + \, 
\sum_{n=1}^{\infty} \, \mathcal{B}_n \, \beta^n \,
\qquad (\beta \to 0^+)\, ,
\label{wcPT_QAHO}
\\
\mathcal{B}_n & = & (-1)^{n+1} \, \frac{4}{\pi^{3/2}} \,
\left( \frac{3}{2} \right)^{n+1/2} \, \Gamma (n+1/2) \, 
\bigl[1 + {\rm O} \left( 1/n \right) \bigr] \,
\qquad (n \to \infty)\, . 
\label{asy_cf_QAHO}
\end{eqnarray}
A comparison of the large-order asymptotics (\ref{e3}) and
(\ref{asy_cf_QAHO}) shows that the two perturbation expansions
(\ref{e2}) and (\ref{wcPT_QAHO}) possess the same rate of divergence if
we choose 
\begin{equation}
\beta \; = \; 40 \, \lambda^2 \, .
\label{beta2lambda}
\end{equation}

In view of these striking similarities between the ground-state
eigenvalues $E_0 (\lambda)$ and $\mathcal{E}_0 (\beta)$ it should be
interesting to investigate what other similarities do exist. In
particular, we are interested in similarities that could provide
evidence that the Pad\'{e} summation of the divergent perturbation
series (\ref{e2}) for the ground-state energy of the
$\mathcal{PT}$-symmetric Hamiltonian (\ref{e1}) converges.

In the case of the ground-state energy shift $\Delta \mathcal{E}_{0}
(\beta)$ of the quartic anharmonic oscillator, which is defined by
\begin{equation}
\mathcal{E}_{0} (\beta) \; = \; 1 \, + \, 
\beta \, \Delta \mathcal{E}_{0} (\beta) \, ,
% \label{HamQAHO}
\end{equation}
it was shown rigorously by Simon (Theorem IV.2.1 of \cite{Si1970}) that
the corresponding perturbation series is a Stieltjes series. This has
some far-reaching consequences. In the case of Stieltjes series,
Pad\'{e} approximants possess a highly developed convergence theory,
as we discuss in detail in Section \ref{Sec:PadEpsStie}. In
particular, the Stieltjes nature of the perturbation series
(\ref{wcPT_QAHO}) guarantees that certain subsequences of the Pad\'{e}
table converge to a uniquely determined Stieltjes function.

Although we cannot prove it rigorously, we believe that for the
$\mathcal{PT}$-symmetric Hamiltonian (\ref{e1}) the ground-state energy
shift $\Delta E_0 (\lambda^2)$ defined by
\begin{equation}
E_0 (\lambda) \; = \; \frac{1}{2} \, + \, \lambda^2 \, 
\Delta E_0 (\lambda^2) \, ,
% \label{}
\end{equation}
in which $E_0$ is considered as a function of $\lambda^2$, is also a
Stieltjes function. This implies that the corresponding perturbation
series is a Stieltjes series.

It is the intention of this article to provide numerical evidence
supporting this conjecture. We do this by comparing Pad\'{e} summations
and Pad\'e predictions (see \cite{We2000} and references therein) of the
perturbation expansions for the ground-state energy shift $\Delta E_0
(\lambda^2)$ and for the analogous ground-state energy shift $\Delta
\mathcal{E}_0 (\beta)$ of the quartic anharmonic oscillator.

Here, one might argue that one should also investigate the summation of
the perturbation expansion for the ground-state energy shift $\Delta E_0
(\lambda^2)$ with the help of the sequence transformations that were
described in Section 7 and 8 of \cite{We1989} and which produced very
good results in the case of the anharmonic oscillators
\cite{WeCiVi1993,We1996b,We1996c}. However, the convergence theory of
these sequence transformations, which in the case of power series also
produce rational approximants, is still very much in its infancy and no
theoretical results concerning the transformation of Stieltjes series
are known so far. Consequently, we would only produce numbers without
gaining any further mathematical insight.

In Section \ref{Sec:PadEpsStie}, we present the
relevant details of Pad\'{e} approximants, the computation of
Pad\'{e} approximants by means of Wynn's recursive epsilon algorithm
\cite{Wy1956}, and we discuss Stieltjes series and their associated Stieltjes
functions. In Section \ref{Sec:SumRes}, we show that the Pad\'{e}
summation of the perturbation expansions for $\Delta E_0 (\lambda^2)$
and $\Delta \mathcal{E}_0 (\beta)$, respectively, produces results of
identical quality if the two coupling constants $\lambda$ and $\beta$
satisfy (\ref{beta2lambda}). In Section \ref{Sec:PadePredict}, we
discuss the prediction of unknown perturbation coefficients with the
help of Wynn's epsilon algorithm, and we show that the coefficients of
the perturbation expansions for either $\Delta E_0 (\lambda^2)$ or
$\Delta \mathcal{E}_0 (\beta)$ can be predicted equally well. Finally,
in Section \ref{Sec:SumConclu} we give a brief summary.

\typeout{==> Section 2} 
\section{Pad\'{e} Approximants, Wynn's Epsilon Algorithm, and Stieltjes 
Series} 
\label{Sec:PadEpsStie} 
  
In recent years, Pad\'{e} approximants have become the standard tool in
theoretical physics to overcome problems with slowly convergent or
divergent power series. Accordingly, there is a vast literature on the
mathematical properties of Pad\'{e} approximants as well as on their
applications in theoretical physics. Any attempt to provide a reasonably
complete biblio\-graphy would be beyond the scope of this article (see
for example the extensive bibliography compiled by Brezinski
\cite{Br1991b}). We just mention that the popularity of Pad\'{e}
approximants in theoretical physics can be traced back to a review by
Baker~\cite{Ba1965}, that the first specialized monograph on Pad\'{e}
approximants is due to Baker \cite{Ba1975}, and that currently the most
complete source of information is the second edition of the monograph by
Baker and Graves-Morris \cite{BaGr1996}. In addition to treatments in
more mathematically oriented books on continued fractions and related
topics \cite{Wal1973,JonThr1980,BowShe1989,LorWaa1992}, Pad\'e
approximants are also discussed in book on mathematical and theoretical
physics, for example in Section 8 of the book by Bender and Orszag
\cite{R18}, or in Part III of a book by Baker on critical phenomena
\cite{Ba1990}. Then, there is a book by Pozzi \cite{Po1994} on the use
of Pad\'{e} approximants in fluid dynamics. Finally, there is even a
monograph \cite{Br1991a} and two articles \cite{Bre1996,Bre2000} on the
history of Pad\'{e} approximants and related topics.
  
A Pad\'{e} approximant $P_{m}^{l} (z)$ to a function $f$ possessing a
(formal) power series expansion
\begin{equation}
f (z) \; = \; \sum_{\nu=0}^{\infty} \, \gamma_{\nu} \, z^{\nu} \, , 
\label{PowSer_f}
\end{equation}
which may converge or diverge, is the ratio of two polynomials $A_l (z)$
and $B_m (z)$ of degrees $l$ and $m$ in $z$ (p.\ 383 of \cite{R18}):
\begin{equation}
P_{m}^{l} (z) \; = \; \frac{A_l (z)}{B_m (z)} \; = \;
\frac{a_0 + a_1 z + a_2 z^2 + \ldots \, a_l z^l}
{1 + b_1 z + b_2 z^2 + \ldots \, b_m z^m} \, .
\label{DefPade}
\end{equation}
An alternative notation for Pad\'{e} approximants, which is
used in the books by Baker and Graves-Morris \cite{Ba1975,BaGr1996}, is
$P_{m}^{l} (z) = [ l / m ]_f (z)$. This notation is usually simplified
further to $P_{m}^{l} (z) = [ l / m ]$ if explicit references to $f$ and
$z$ are not necessary.
  
The coefficients $a_{\lambda}$ and $b_{\mu}$ of the polynomials $A_l
(z)$ and $B_m (z)$ in (\ref{DefPade}) are chosen in such a way that the
Taylor expansion of $f (z)$ and of its Pad\'{e} approximant agree as far
as possible:
\begin{equation}
f (z) - P_{m}^{l} (z) \; = \; \mathrm{O} \left( z^{l+m+1} \right) \,
\qquad (z \to 0) \, .
\label{OestPA}
\end{equation}
This asymptotic error estimate leads to a system of linear equations by
means of which the coefficients $a_0$, $a_1$, \ldots, $a_l$ and $b_1$,
$b_2$, \ldots, $b_m$ in (\ref{DefPade}) can be computed
\cite{Ba1975,BaGr1996}. Moreover, several algorithms are known which 
permit a recursive computation of Pad\'e approximants. A discussion of
the merits and weaknesses of the various computational schemes can for
instance be found in Section II.3 of the book by Cuyt and Wuytack
\cite{CuyWuy1987}.
  
Probably, the best known recursive algorithm for Pad\'{e} approximants is
Wynn's epsilon algorithm \cite{Wy1956}:
\begin{subequations}
\label{eps_al}
\begin{eqnarray}
\varepsilon_{-1}^{(n)} & \; = \; & 0 \, ,
\qquad \varepsilon_0^{(n)} \; = \; s_n \, ,
\qquad n \in \mathbb{N}_0 \, , \\
\varepsilon_{k+1}^{(n)} & \; = \; & \varepsilon_{k-1}^{(n+1)} \, + \,
1 / [\varepsilon_{k}^{(n+1)} - \varepsilon_{k}^{(n)} ] \, ,
\qquad k, n \in \mathbb{N}_0 \, .
\end{eqnarray}           
\end{subequations}
A compact FORTRAN program for the epsilon algorithm as well as the
underlying computational algorithm is described in Section 4.3 of
\cite{We1989}.
  
If the input data $\varepsilon_0^{(n)} = s_n$ of Wynn's epsilon
algorithm are the partial sums
\begin{equation}
f_n (z) \; = \; 
\sum_{\nu=0}^{n} \, \gamma_{\nu} \, z^{\nu}
\label{ParSumSer_f}
\end{equation} 
of the formal power series (\ref{PowSer_f}) according to
$\varepsilon_0^{(n)} = f_n (z)$, then the elements
$\varepsilon_{2k}^{(n)}$ with \emph{even} subscripts are Pad\'{e}
approximants to $f (z)$ according to \cite{Wy1956}
\begin{equation}
\varepsilon_{2k}^{(n)} \; = \; P_{k}^{k+n} (z) \, .
\label{Eps_Pade}
\end{equation} 
The elements $\varepsilon_{2k+1}^{(n)}$ with \emph{odd} subscripts are
only auxiliary quantities, which diverge if the whole transformation
process converges.
 
The epsilon algorithm is a useful numerical algorithm that is
applied successfully in a large variety of different fields. 
Accordingly, there is an extensive literature dealing with it. A
fairly complete coverage of the older literature can be found in a book
by Brezinski \cite{Bre1977}. It may be interesting to note that the
epsilon algorithm is not restricted to scalar sequences but can be
generalized to cover vector sequences. A recent review
of these developments can be found in \cite{GMRobSal2000}.
   
If one tries to sum a divergent power series by converting its partial
sums (\ref{ParSumSer_f}) to Pad\'e approximants, it is usually a good
idea to use \emph{diagonal} Pad\'{e} approximants, whose
numerator and denominator polynomials have equal degrees. If this
is not possible one should use Pad\'{e} approximants with numerator and
denominator polynomials whose degrees differ as little as possible. If
we use the epsilon algorithm for the computation of the Pad\'e
approximants, then (\ref{Eps_Pade}) implies that we should use the
elements of following staircase sequence in the Pad\'e table as
approximations to $f (z)$ (see Eq.\ (4.3-7) of \cite{We1989}):
\begin{equation}
P_{0}^{0} (z), P_{0}^{1} (z), P_{1}^{1} (z), \ldots , P_{\nu}^{\nu} (z),
P_{\nu}^{\nu+1} (z), P_{\nu+1}^{\nu+1} (z), \ldots \, .
% \label{}
\end{equation}
This staircase sequence exploits the available information optimally if
the partial sums $f_n (z)$ with $n \ge 0$ are computed successively and
if, after the computation of each new partial sum, the element of the
epsilon table with the highest possible {\em even\/} subscript is
computed. With the help of the notation $\Ent{x}$ for the integral part
of $x$, this staircase sequence can be written compactly as follows:
\begin{equation}
\varepsilon_{2 \Ent{n/2}}^{(n - 2 \Ent{n/2})} \; = \;
P_{\Ent{n/2}}^{n-\Ent{n/2}} (z) \,
\qquad (n = 0, 1, 2, \ldots) \, .  
\label{EpsPadeAppSeq}
\end{equation}
  
As remarked above, Pad\'{e} approximants are now used almost routinely
to overcome problems with slowly convergent or divergent power
series. Hence, their practical usefulness is beyond question. However,
from a theoretical point of view, the situation is not so good. So far,
a completely satisfactory \emph{general} convergence theory of Pad\'{e}
approximants for essentially arbitrary power series does not exist.

Nevertheless, there is a special class of series, the so-called
\emph{Stieltjes} series, which possess a highly developed and elegant
convergence theory. In this section we will only discuss those
properties of Stieltjes series and Stieltjes functions that are needed
to provide numerical evidence that the perturbation expansion for the
ground-state energy shift $\Delta E_0 (\lambda^2)$ of the complex
$\mathcal{PT}$-symmetric Hamiltonian (\ref{e1}), considered as a
function of $\lambda^2$, is a Stieltjes series. Detailed discussions of
the properties of Stieltjes series and their special role in the theory
of Pad\'{e} approximants can be found in Section 8.6 of
\cite{R18} or in Section 5 of \cite{BaGr1996}.

A function $F (z)$ with $z \in \mathbb{C}$ is called a \emph{Stieltjes}
function if it can be expressed as a Stieltjes integral according to
\begin{equation}
F (z) \; = \; \int_{0}^{\infty} \, 
\frac{\mathrm{d} \Phi (t)} {1+zt} \,
\qquad [\vert \arg (z) \vert < \pi] \, .
% \label{StieFun}
\end{equation}
Here, $\Phi (t)$ is a bounded, nondecreasing function taking infinitely
many different values on $0 \le t < \infty$. Moreover, the moment integrals
\begin{equation}
\mu_n \; = \; \int_{0}^{\infty} \, t^n \, \mathrm{d} \Phi (t) \, 
\qquad (n \in \mathbb{N}_0)\,
% \label{StieMom}
\end{equation}
must be real and finite for all finite values of $n$. A Stieltjes
function can be expressed by its corresponding
\emph{Stieltjes} series:
\begin{equation}
F (z) \; = \; 
\sum_{\nu=0}^{\infty} \, (-1)^{\nu} \, \mu_{\nu} \, z^{\nu} \, .
\label{StieSer}
\end{equation}
Whether this series converges or diverges depends on the behavior of the 
Stieltjes moments $\mu_n$ as $n \to \infty$.

In a typical Stieltjes summation problem, as it occurs in
the context of divergent perturbation expansion, only the numerical
values of a finite number of Stieltjes moments $\mu_n$ are known. Thus,
one has to find a way of constructing an approximation to the unknown
Stieltjes function $F (z)$ from a finite string of moments.

Of course, one would also like to have some theoretical evidence that $F
(z)$ exists and is uniquely determined by the Stieltjes moments $\{
\mu_n \}_{n=0}^{\infty}$. Many necessary and sufficient conditions
that guarantee this are known in literature. Unfortunately, it is by
no means easy to apply them. For example, a necessary condition that the
series (\ref{StieSer}) is indeed a Stieltjes series is that the Hankel
determinants
\begin{equation}
D (m, n) \; = \; 
\begin{vmatrix}
\mu_{m} & \mu_{m+1} & \dots & \mu_{m+n} \\
\mu_{m+1} & \mu_{m+2} & \dots & \mu_{m+n+1} \\
\vdots & \vdots & \ddots & \vdots \\
\mu_{m+n} & \mu_{m+n+1} & \dots & \mu_{m+2n} \\
\end{vmatrix}
% \label{}
\end{equation}
are positive for all $m, n \ge 0$ (see Theorem 5.1.2\ on
p.\ 197 of \cite{BaGr1996}). 

Then, there is a sufficient criterion, the so-called Carleman condition
(see p.\ 410 of \cite{R18} or pp.\ 239 - 240 of
\cite{BaGr1996}), which requires that the series $\sum_{j=1}^{\infty}
(\mu_j)^{-1/(2j)}$ diverges and thus limits the admissible growth of the
moments $\mu_n$ as $n \to \infty$ \cite{GraGre1978}. If the condition
$D (m, n)>0$ on the determinants as well as the Carleman
condition are both satisfied, then the Pad\'{e} approximants $P_{m}^{m+j}
(z)$ constructed from the partial sums of the moment expansion
(\ref{StieSer}) converge for every $j \ge - 1$ to the corresponding
Stieltjes function $F (z)$ as $m \to \infty$ (see for example Theorem
5.5.1 on p.\ 240 of \cite{BaGr1996}).

If only a finite number of moments are known, it is impossible to prove
that $D (m, n) > 0$ holds for all $m, n \ge 0$, and it is also not
possible to prove rigorously that the Carleman condition is satisfied,
although we would like to emphasize that the large-order formula
(\ref{e3}), which was verified numerically in \cite{R13}, is in
agreement with the Carleman condition.

Therefore, we prefer an indirect approach in order to provide evidence
that the perturbation series for the energy shift $\Delta E_0
(\lambda^2)$ of the $\mathcal{PT}$-symmetric Hamiltonian (\ref{e1}) is
indeed a Stieltjes series. For that purpose, let us assume that the moment
expansion (\ref{StieSer}), whose Stieltjes nature we want to establish,
is a Stieltjes series. Pad\'{e} approximants to Stieltjes series
possess a highly developed convergence theory, and many conditions and
inequalities are known that Pad\'{e} approximants to a Stieltjes series
must satisfy. For example, Pad\'{e} approximants constructed from the
partial sums
\begin{equation}
F_n (z) \; = \; 
\sum_{\nu=0}^{n} \, (-1)^{\nu} \, \mu_{\nu} \, z^{\nu}
\label{ParSumStieSer}
\end{equation}
of the moment expansion (\ref{StieSer}) for a Stieltjes
function $F (z)$ satisfy for $z > 0$ the following inequalities (Theorem
15.2\ on p.\ 215 of
\cite{Ba1975}):
\begin{eqnarray}
\label{PadIneq_1}
(-1)^{j+1} \, \left\{ P_{m+1}^{m+j+1} (z) \, - \, 
P_{m}^{m+j} (z) \right\} \; \ge & 0 \, , \\
\label{PadIneq_2}
(-1)^{j+1} \, \left\{ P_{m}^{m+j} (z) \, - \,
P_{m-1}^{m+j+1} (z) \right\} \; \ge & 0 \, , \\
P_{m}^{m} (z) \; \ge \; F (z) \; \ge \; P_{m}^{m-1} (z) \, . &  
\label{PadIneq_3}
\end{eqnarray}

It follows from inequality (\ref{PadIneq_1}) that the Pad\'{e} sequence
$\bigl\{ P_{m}^{m+j} (z) \bigr\}_{m=0}^{\infty}$ is \emph{increasing} for $z>0$
if $j$ is \emph{odd}, and it is \emph{decreasing} if $j$ is \emph{even}.
Moreover, if we set $j = - 1$ in (\ref{PadIneq_2})
and replace $m$ by $m+1$, we obtain the inequality
\begin{equation}
P_{m+1}^{m} (z) \; \ge \; P_{m}^{m+1} (z) \,
\qquad (m \in \mathbb{N}_0)\, .
\label{PadIneq_4}
\end{equation}
Thus, if we use Wynn's epsilon algorithm (\ref{eps_al}) to convert the
partial sums (\ref{ParSumStieSer}) to Pad\'{e} approximants and choose
the approximants to $F (z)$ according to (\ref{EpsPadeAppSeq}), then it
follows from inequalities (\ref{PadIneq_3}) and (\ref{PadIneq_4}) that
these Pad\'{e} approximants satisfy the following inequality if the
moment expansion (\ref{StieSer}) for $F (z)$ is a Stieltjes series:
\begin{equation}
P_{m}^{m+1} (z) \; \le \; F (z) \; \le \; P_{m+1}^{m+1} (z) \,
\qquad (z > 0 \, , ~ m \in \mathbb{N}_0)\, .
\label{EpsPadeIneq}
\end{equation}
Thus, the approximants (\ref{EpsPadeAppSeq}) produced by Wynn's epsilon
algorithm yield for $z > 0$ two nesting sequences $P_{m}^{m+1} (z) =
\varepsilon_{2m}^{(1)}$ and $P_{m+1}^{m+1} (z) = 
\varepsilon_{2m+2}^{(0)}$ of lower and upper bounds to $F (z)$ if the 
the moment expansion (\ref{StieSer}) is a Stieltjes series. 

If only the numerical values of a finite number of Stieltjes moments are
available, then it is of course not possible to prove rigorously that
the series under consideration is a Stieltjes series. Nevertheless, we
can provide considerable evidence that this hypothesis is true if
inequality (\ref{EpsPadeIneq}) is valid in all cases that can be checked.

\typeout{==> Section 3} 
\section{Summation Results} 
\label{Sec:SumRes}

In this section we want to show that the Pad\'{e} summation of the
perturbation expansion
\begin{equation}
\Delta E_0 (\lambda^2) \; = \; 
\sum_{\nu=0}^{\infty} \, b_{\nu+1} \, \lambda^{2 \nu}
\label{PerSerDelE0PT}
\end{equation}
for the ground-state energy shift of the of the $\mathcal{PT}$-symmetric
Hamiltonian (\ref{e1}) and of the perturbation expansion 
\begin{equation}
\Delta \mathcal{E}_{0} (\beta) \; = \; 
\sum_{\nu=0}^{\infty} \, \mathcal{B}_{\nu+1} \, \beta^{\nu} 
\label{PerSerDelE0QAHO}
\end{equation}
for the ground-state energy shift of the quartic anharmonic oscillator
yield results of virtually identical quality if the two coupling
constants $\lambda$ and $\beta$ satisfy (\ref{beta2lambda}). Moreover,
we want to demonstrate numerically that inequality (\ref{EpsPadeIneq}),
which is satisfied in the case of the Pad\'e summation a Stieltjes
series, is apparently also satisfied. In our summation calculations, we
used all coefficients $b_{\nu}$ and $\mathcal{B}_{\nu}$ with $0 \le \nu
\le 193$ which had been computed recursively.

In this article, we compute all Pad\'{e} approximants with the help of
Wynn's epsilon algorithm (\ref{eps_al}).  Thus, the partial sums
\begin{equation}
s_n (\lambda^2) \; = \; 
\sum_{\nu=0}^{n} \, b_{\nu+1} \, \lambda^{2 \nu}
\label{ParSumPT}
\end{equation}
and 
\begin{equation}
\sigma_n (\beta) \; = \; 
\sum_{\nu=0}^{n} \, \mathcal{B}_{\nu+1} \, \beta^{\nu} 
\label{ParSumQAHO}
\end{equation}
are used as input data for Wynn's epsilon algorithm according to
$\varepsilon_{0}^{(n)} = s_n (\lambda^2)$ or $\varepsilon_{0}^{(n)} =
\sigma_n (\beta)$, respectively, and the approximations to the energy
shifts are chosen according to (\ref{EpsPadeAppSeq}).

\begin{table}[htb] 
\caption{\label{Table_2} Pad\'e summation of the perturbation 
expansions (\protect\ref{PerSerDelE0PT}) and
(\protect\ref{PerSerDelE0QAHO}) for the ground-state energy shifts
$\Delta E_0 (\lambda^2)$ and $\Delta
\mathcal{E}_{0} (\beta)$ with $\lambda = 1/7$ and $\beta = 40/49$,
respectively.}
\begin{tabular} {lrrrr} \\
[1\jot] \hline \hline \rule{0pt}{4\jot}%
$n$ & \multicolumn{1}{c}{$s_n (\lambda^2)$} 
& \multicolumn{1}{c}{$\sigma_n (\beta)$} 
& \multicolumn{1}{c}{$P_{\Ent{n/2}}^{n-\Ent{n/2}} (\lambda^2)$} 
& \multicolumn{1}{c}{$P_{\Ent{n/2}}^{n-\Ent{n/2}} (\beta)$} \\ 
[1\jot] \hline \rule{0pt}{4\jot}%
0   & $ 0.110 \cdot 10^{002}$ & $ 0.750 \cdot 10^{000}$ & 
$11.00~000~000~000~000$ & $ 0.750~000~000~000~000$ \\
1   & $-0.798 \cdot 10^{001}$ & $-0.321 \cdot 10^{000}$ & 
$-7.97~959~183~673~469$ & $-0.321~428~571~428~571$ \\
2   & $ 0.582 \cdot 10^{002}$ & $ 0.315 \cdot 10^{001}$ & 
$ 6.76~871~520~405~468$ & $ 0.497~075~017~205~781$ \\
3   & $-0.269 \cdot 10^{003}$ & $-0.133 \cdot 10^{002}$ & 
$ 3.14~452~476~154~168$ & $ 0.283~471~705~042~096$ \\
4   & $ 0.177 \cdot 10^{004}$ & $ 0.861 \cdot 10^{002}$ & 
$ 5.92~770~890~838~469$ & $ 0.444~962~648~249~413$ \\
5   & $-0.134 \cdot 10^{005}$ & $-0.639 \cdot 10^{003}$ & 
$ 4.84~920~642~167~536$ & $ 0.379~736~282~027~717$ \\
[1\jot] \hline \rule{0pt}{4\jot}%
50  & $ 0.153 \cdot 10^{072}$ & $ 0.684 \cdot 10^{070}$ & 
$ 5.52~416~958~165~793$ & $ 0.419~249~574~461~710$ \\
51  & $-0.964 \cdot 10^{073}$ & $-0.432 \cdot 10^{072}$ & 
$ 5.52~416~451~428~038$ & $ 0.419~249~241~261~250$ \\
52  & $ 0.620 \cdot 10^{075}$ & $ 0.278 \cdot 10^{074}$ & 
$ 5.52~416~888~260~688$ & $ 0.419~249~527~748~761$ \\
53  & $-0.407 \cdot 10^{077}$ & $-0.182 \cdot 10^{076}$ & 
$ 5.52~416~531~636~255$ & $ 0.419~249~293~076~390$ \\
54  & $ 0.272 \cdot 10^{079}$ & $ 0.122 \cdot 10^{078}$ & 
$ 5.52~416~839~738~891$ & $ 0.419~249~495~310~895$ \\
[1\jot] \hline \rule{0pt}{4\jot}%
101 & $-0.210 \cdot 10^{172}$ & $-0.939 \cdot 10^{170}$ & 
$ 5.52~416~721~141~847$ & $ 0.419~249~415~925~473$ \\
102 & $ 0.263 \cdot 10^{174}$ & $ 0.118 \cdot 10^{173}$ & 
$ 5.52~416~721~422~990$ & $ 0.419~249~416~112~202$ \\
103 & $-0.334 \cdot 10^{176}$ & $-0.149 \cdot 10^{175}$ & 
$ 5.52~416~721~178~460$ & $ 0.419~249~415~949~529$ \\
104 & $ 0.427 \cdot 10^{178}$ & $ 0.191 \cdot 10^{177}$ & 
$ 5.52~416~721~397~212$ & $ 0.419~249~416~094~862$ \\
105 & $-0.552 \cdot 10^{180}$ & $-0.247 \cdot 10^{179}$ & 
$ 5.52~416~721~206~667$ & $ 0.419~249~415~968~069$ \\
[1\jot] \hline \rule{0pt}{4\jot}%
150 & $ 0.318 \cdot 10^{279}$ & $ 0.142 \cdot 10^{278}$ & 
$ 5.52~416~721~306~531$ & $ 0.419~249~416~033~824$ \\
151 & $-0.590 \cdot 10^{281}$ & $-0.264 \cdot 10^{280}$ & 
$ 5.52~416~721~305~477$ & $ 0.419~249~416~033~119$ \\
152 & $ 0.110 \cdot 10^{284}$ & $ 0.493 \cdot 10^{282}$ & 
$ 5.52~416~721~306~436$ & $ 0.419~249~416~033~760$ \\
153 & $-0.207 \cdot 10^{286}$ & $-0.928 \cdot 10^{284}$ & 
$ 5.52~416~721~305~579$ & $ 0.419~249~416~033~187$ \\
154 & $ 0.392 \cdot 10^{288}$ & $ 0.175 \cdot 10^{287}$ & 
$ 5.52~416~721~306~359$ & $ 0.419~249~416~033~708$ \\
[1\jot] \hline \rule{0pt}{4\jot}%
188 & $ 0.371 \cdot 10^{367}$ & $ 0.166 \cdot 10^{366}$ & 
$ 5.52~416~721~306~035$ & $ 0.419~249~416~033~489$ \\
189 & $-0.860 \cdot 10^{369}$ & $-0.385 \cdot 10^{368}$ & 
$ 5.52~416~721~306~009$ & $ 0.419~249~416~033~472$ \\
190 & $ 0.201 \cdot 10^{372}$ & $ 0.898 \cdot 10^{370}$ & 
$ 5.52~416~721~306~033$ & $ 0.419~249~416~033~488$ \\
191 & $-0.471 \cdot 10^{374}$ & $-0.211 \cdot 10^{373}$ & 
$ 5.52~416~721~306~011$ & $ 0.419~249~416~033~474$ \\
192 & $ 0.111 \cdot 10^{377}$ & $ 0.496 \cdot 10^{375}$ & 
$ 5.52~416~721~306~031$ & $ 0.419~249~416~033~487$ \\
[1\jot] \hline \hline \rule{0pt}{4\jot}%
\end{tabular}
\end{table}

In Table \ref{Table_2} we present illustrative results of the Pad\'{e}
summation of the perturbation expansion (\ref{PerSerDelE0PT}) for the
ground-state energy shift $\Delta E_0 (\lambda^2)$ with $\lambda = 1/7$
and of the perturbation expansion (\ref{PerSerDelE0QAHO}) for the
ground-state energy shift $\Delta \mathcal{E}_{0} (\beta)$ with $\beta =
40/49$. Thus, the two coupling constants $\lambda$ and $\beta$ satisfy
(\ref{beta2lambda}). This implies that the two perturbation expansions,
whose partial sums are displayed in columns 2 and 3, should show the
same rate of divergence.

Here, we must remember that the larger-order estimates (\ref{e3}) and
(\ref{asy_cf_QAHO}), respectively, imply that the partial sums
(\ref{ParSumPT}) for the ground-state energy shift $\Delta E_0
(\lambda^2)$ of the $\mathcal{PT}$-symmetric Hamiltonian (\ref{e1}) are
for all $n \ge 0$ at least one order of magnitude greater than the
partial sums (\ref{ParSumQAHO}) for the ground-state energy shift
$\Delta \mathcal{E}_{0} (\beta)$ of the quartic anharmonic oscillator.
Otherwise, the observed rates of divergence in columns 2 and 3 are
virtually identical.

Moreover, the Pad\'e approximants in columns 4 and 5 apparently satisfy the
inequality (\ref{EpsPadeIneq}), which holds if the series to be
transformed is a Stieltjes series. If the index $n$ in column 1 is even
($n = 2m$), then the diagonal Pad\'e approximants $P_{m}^{m} (\lambda^2)$
and $P_{m}^{m} (\beta)$ provide upper bounds that strictly decrease
with increasing $m$ and, if $n$ is odd ($n = 2m+1$), then the Pad\'e
approximants $P_{m}^{m+1} (\lambda^2)$ and $P_{m}^{m+1} (\beta)$ provide
lower bounds that strictly increase with increasing $m$.

We have done analogous summation calculations also for many other values of
the coupling constants $\lambda$ and $\beta$. Of course, the performance
of the Pad\'e summations depend very much on the size of the coupling
constants. For smaller values of $\lambda$ and the corresponding
$\beta$, convergence is better than in Table \ref{Table_2}, whereas for
$\lambda = 1$ and $\beta = 40$ only the first digit of the summation
results stabilize. For larger values of $\lambda$ and $\beta$, Pad\'e
summation produces only relatively crude upper and lower
bounds. However, to emphasize that the typical qualitative
features of the summation results in Table \ref{Table_2} -- the same
rate of divergence of two perturbation series and the occurrence of
strictly decreasing upper bounds $P_{m}^{m}$ and strictly increasing
lower bounds $P_{m}^{m+1}$ -- were consistently observed in all cases
considered.

Thus, Wynn's epsilon algorithm is apparently unable to detect any
substantial difference between the perturbation series
(\ref{PerSerDelE0QAHO}) for $\Delta \mathcal{E}_{0} (\beta)$, whose
Stieltjes nature was established rigorously by Simon (Theorem IV.2.1 of
\cite{Si1970}), and the perturbation series (\ref{PerSerDelE0PT}) for
$\Delta E_0 (\lambda^2)$, whose Stieltjes nature we conjecture.

\typeout{==> Section 4} 
\section{Pad\'{e} Predictions} 
\label{Sec:PadePredict}

As shown by countless articles from all branches of physics, Pad\'{e}
approximants have become the standard tool to overcome problems with
slowly convergent or divergent power series. However, Pad\'{e}
approximants have other useful features that are not as well known yet. For
example, Pad\'{e} approximants can be used to make predictions for
higher-order series coefficients that were not used for 
the construction of the approximant.

On a heuristic level the prediction capability of Pad\'{e}
approximants, which was apparently first noted and used by Gilewicz
\cite{Gil1973}, can be explained quite easily. Let us assume that the 
partial sums (\ref{ParSumSer_f}) of the power series for some function
$f (z)$ are to be converted to Pad\'{e} approximants. Then, the
\emph{accuracy-through-order} relationship (\ref{OestPA}) implies that a
Pad\'{e} approximant $P_{m}^{l} (z)$ to $f (z)$ can be expressed as the
partial sum $f_{l+m} (z)$ from which it was constructed plus a term
$z^{l+m+1} \mathcal{R}_{m}^{l} (z)$, which was generated by the
transformation of the partial sum to the rational approximant:
\begin{equation}
P_{m}^{l} (z) \; = \; \sum_{\nu=0}^{l+m} \, \gamma_{\nu} \, z^{\nu} 
\, + \, z^{l+m+1} \, \mathcal{R}_{m}^{l} (z)
\; = \; f_{l+m} (z) \, + \, z^{l+m+1} \, \mathcal{R}_{m}^{l} (z) \, .
\label{TranTermPA}
\end{equation}
Similarly, the power series (\ref{PowSer_f}) can be expressed as
follows:
\begin{equation}
f (z) \; = \; \sum_{\nu=0}^{l+m} \, \gamma_{\nu} \, z^{\nu} 
\, + \, z^{l+m+1} \, 
\sum_{\nu=0}^{\infty} \, \gamma_{l+m+\nu+1} \, z^{\nu}
\; = \; f_{l+m} (z) \, + \, z^{l+m+1} \, \mathcal{F}_{l+m+1} (z) \, .
% \label{}
\end{equation}
Let us now assume that the indices $l$ and $m$ so large
that the Pad\'{e} approximant $P_{m}^{l} (z)$ provides a sufficiently
accurate approximation to $f (z)$. Then, the Pad\'{e} transformation
term $\mathcal{R}_{m}^{l} (z)$ must also provide a sufficiently accurate
approximation to the truncation error $\mathcal{F}_{l+m+1} (z)$ of the
power series. In general, we have no reason to assume that
\begin{equation}
\mathcal{R}_{m}^{l} (z) \; = \; \mathcal{F}_{l+m+1} (z)
% \label{}
\end{equation}
might hold \emph{exactly} for finite values of $l$ and
$m$. Consequently, Taylor expansions of $\mathcal{R}_{m}^{l} (z)$ and
$\mathcal{F}_{l+m+1} (z)$, respectively, will in general produce
different results. Nevertheless, the \emph{leading} coefficients of the
Taylor expansion for $\mathcal{R}_{m}^{l} (z)$ should in such a case
provide sufficiently accurate approximations to the corresponding
coefficients of the Taylor series for $\mathcal{F}_{l+m+1} (z)$.

It is important to note that this prediction capability of Pad\'e
approximants does not depend on the convergence of the power series
expansions for $\mathcal{R}_{m}^{l} (z)$ and $\mathcal{F}_{l+m+1} (z)$,
respectively, which was used implicitly in our heuristic reasoning given
above. Pad\'{e} approximants are able to make predictions about series
coefficients even if the power series (\ref{PowSer_f}) for $f$ as well
as the power series expansions for $\mathcal{R}_{l}^{m}$ and
$\mathcal{F}_{l+m+1} (z)$ are only asymptotic as $z \to 0$. This fact
explains why the prediction capability of Pad\'{e} approximants can be
so very useful in the case of violently divergent perturbation
expansions (see \cite{We2000,SaLiSt1993,SaLi1994,%
SaLiSt1994,SaElKa1995,SaLiSt1995_97,ElGaKaSa1996a,ElGaKaSa1996b,%
We1997,SaAbYu1997,JaJoSa1997,ElJaJoKaSa1998,ElStChMiSp1998,StEl1998,%
ChElSt1999a,ChElSt1999b,ChElSt2000a,ChElSt2000b,%
JeBeWeSo1999,JeWeSo2000} and references therein).

Theoretically, very little is known about the prediction of series
coefficients that were not used for the construction of the Pad\'e
approximant. A notable exception are again Stieltjes series for which
inequalities are known. 
 
Let us assume that the partial sums (\ref{ParSumStieSer}) of the moment
expansion for some Stieltjes function $F (z)$ are to be converted to
Pad\'e approximants. Such a Pad\'e approximant $P_{m}^{l} (z)$ possesses
the following power series expansion:
\begin{equation}
P_{m}^{l} (z) \; = \; 
\sum_{\nu=0}^{\infty} \, (-1)^{\nu} \, \mu^{[l/m]}_{\nu} \, z^{\nu} \, .
\label{PowSerStiePade}
\end{equation}
In Theorem 5.2.7\ on p.\ 220 of \cite{BaGr1996} it was shown that for
all $n \in \mathbb{N}_0$ and for all $l \ge m - 1$ the coefficients
$\mu^{[l/m]}_n$ in (\ref{PowSerStiePade}) are bounded in magnitude by
the Stieltjes moments $\mu_n$ in (\ref{StieSer}) according to
\begin{equation}
0 \; \le \; \mu^{[l/m]}_n \; \le \; \mu_n \, .
\label{PadeMom_le_StieMom}
\end{equation}
 
This inequality can be used to analyze the Stieltjes nature of a moment
expansion of the type of (\ref{StieSer}). With the help of computer
algebra systems like Maple or Mathematica it is possible to construct
Pad\'e approximants $P_{m}^{l} (z)$ in an unspecified symbolic variable
$z$, and this can even be done free of rounding errors if the
coefficients of the series to be transformed are exact integers like the
coefficients $b_n$ in (\ref{e2}) or exact rational numbers like the
coefficients $\mathcal{B}_n$ in (\ref{wcPT_QAHO}). In the next step, a
the leading part of a power series expansion of the Pad\'e approximant
must be constructed. The resulting series coefficients can then be
compared with the corresponding coefficients of the moment expansion.
 
Again, this poses no principal problems for computer algebra systems
like Maple and Mathematica. However, the accuracy-through-order
relationship (\ref{OestPA}) implies that inequality
(\ref{PadeMom_le_StieMom}), which is to be checked, is by default
satisfied for all indices $n \le l + m$, and only for $n \ge l + m + 1$
we obtain useful information about the Stieltjes nature of the moment
expansion, from which the Pad\'e approximant was constructed. Thus, if
$l + m$ becomes large, the brute-force approach based on computer
algebra systems becomes very demanding both with respect to computer
time and memory because it requires both the symbolic construction of
complicated Pad\'e approximants and also symbolic differentiations of
very high orders.
 
These computational problems can be simplified considerably with the
help of a recently derived recursive scheme (Section 3 of
\cite{We2000}) that permits a direct calculation of the transformation
term $\mathcal{R}_{m}^{l} (z)$ in (\ref{TranTermPA}) if the
corresponding Pad\'e approximant $P_{m}^{l} (z)$ can be computed with
the help of Wynn's epsilon algorithm according to (\ref{Eps_Pade});
that is, for Pad\'e approximants of the type $P_{k}^{k+n} (z)$ with $k, n
\in \mathbb{N}_0$.
 
It follows from the accuracy-through-order relationship (\ref{OestPA})
in combination with (\ref{Eps_Pade}) that $\varepsilon_{2k}^{(n)}$ can be
expressed as follows if the partial sums (\ref{ParSumSer_f}) of the
power series for some function $f (z)$ are used as input data for Wynn's 
epsilon algorithm (\ref{eps_al}):
\begin{equation}
\varepsilon_{2k}^{(n)} \; = \; f_{n+2k} (z) \, + \, z^{n+2k+1} \, 
\varphi_{2k}^{(n)} (z) \, .
\label{EpsRem_phi}
\end{equation}
The quantities $\varphi_{2k}^{(n)} (z)$ can be computed with the help of
the recursive scheme  in Eq.\ (3.15) of \cite{We2000}, which uses
the coefficients $\gamma_n$ of the power series (\ref{PowSer_f}) for $f
(z)$ as input data, as follows:
\begin{subequations}
\label{rec_phi}
\begin{eqnarray}
\varphi_{0}^{(n)} (z) & = & 0 \, , \qquad n \in \mathbb{N}_0 \, ,
% \label{}
\\
\varphi_{2}^{(n)} (z) & = & \frac
{\bigl[ \gamma_{n+2} \bigr]^2}{\gamma_{n+1} - \gamma_{n+2} z} \, ,
\qquad n \in \mathbb{N}_0 \, ,
% \label{}
\\
\varphi_{2k+2}^{(n)} (z) & = & \varphi_{2k}^{(n+2)} (z) \, + \,
\frac{\alpha_{2k+2}^{(n)} (z)}{\beta_{2k+2}^{(n)} (z)} \, ,
\qquad k \in \mathbb{N} \, , \quad n \in \mathbb{N}_0 \, ,
% \label{}
\\
\alpha_{2k+2}^{(n)} (z) & = & 
\frac{\gamma_{n+2k+2} + \delta \varphi_{2k}^{(n+1)} (z)}
{\gamma_{n+2k+1} + \delta \varphi_{2k}^{(n)} (z)} \, - \,
\frac{\gamma_{n+2k+2} + \delta \varphi_{2k}^{(n+1)} (z)}
{\gamma_{n+2k+1} + z \varphi_{2k}^{(n+1)} (z) - 
\varphi_{2k-2}^{(n+2)} (z)} \, ,
% \label{}
\\
\beta_{2k+2}^{(n)} (z) & = & 
\frac{1}{\gamma_{n+2k+2} + \delta \varphi_{2k}^{(n+1)} (z)} \, - \,
\frac{z}{\gamma_{n+2k+1} + \delta \varphi_{2k}^{(n)} (z)} \nonumber \\ 
& & \, + \,
\frac{z}{\gamma_{n+2k+1} + z \varphi_{2k}^{(n+1)} (z) - 
\varphi_{2k-2}^{(n+2)} (z)} \, ,
% \label{}
\end{eqnarray}
\end{subequations}
where
\begin{equation}
\delta \varphi_{2k}^{(n)} (z) \; = \; 
z \varphi_{2k}^{(n+1)} (z) \, - \, \varphi_{2k}^{(n)} (z) \, .
% \label{}
\end{equation}
The rational function $\varphi_{2k}^{(n)} (z)$ can be expressed as a
power series in $z$ according to
\begin{equation}
\varphi_{2k}^{(n)} (z) \; = \; g_{0}^{(n, 2k)} \, + \, 
g_{1}^{(n, 2k)} \, z \, + \, g_{2}^{(n, 2k)} \, z^2 \, + \, \ldots
\, + \, g_{\nu}^{(n, 2k)} \, z^{\nu} \, + \, \ldots \, . 
\label{PowSerPhi}
\end{equation}
The coefficients $g_{\nu}^{(n, 2k)}$ of this series expansion can be used to
predict the coefficients $\gamma_{n+2k+\nu+1}$ with $\nu \ge 0$
of the power series expansion (\ref{PowSer_f}) for $f (z)$ that were
not used for the construction of either $\varepsilon_{2k}^{(n)}$ or
$\varphi_{2k}^{(n)} (z)$.

Thus, we can compute the rational function $\varphi_{2k}^{(n)}$ with the 
help of the recursive scheme (\ref{rec_phi}) in the case of the
perturbation series (\ref{PerSerDelE0PT}) for $\Delta E_0 (\lambda^2)$
and (\ref{PerSerDelE0QAHO}) for $\Delta \mathcal{E}_{0} (\beta)$. In the 
case of the $\mathcal{PT}$-symmetric perturbation series we obtain
the expansion
\begin{equation}
\varphi_{2k}^{(n)} (\lambda^2) \; = \; b_{0}^{(n, 2k)} \, + \,
b_{1}^{(n, 2k)} \, \lambda^2 \, + \, b_{2}^{(n, 2k)} \, \lambda^4 
\, + \, \ldots \, + \, b_{\nu}^{(n, 2k)} \, \lambda^{2\nu} 
\, + \, \ldots \, ,
\label{PhiSerPT}
\end{equation}
and in the case of the perturbation series for the quartic anharmonic
oscillator we obtain
\begin{equation}
\varphi_{2k}^{(n)} (\beta) \; = \; \mathcal{B}_{0}^{(n, 2k)} \, + \,
\mathcal{B}_{1}^{(n, 2k)} \, \beta \, + \, 
\mathcal{B}_{2}^{(n, 2k)} \, \beta^2 
\, + \, \ldots \, + \, \mathcal{B}_{\nu}^{(n, 2k)} \, \beta^{\nu} 
\, + \, \ldots \, .
\label{PhiSerX4}
\end{equation}
The coefficients $b_{\nu}^{(n, 2k)}$ and $\mathcal{B}_{\nu}^{(n, 2k)}$
with $\nu \ge 0$ can be used to obtain predictions for the coefficients
$b_{n+2k+\nu+2}$ and $\mathcal{B}_{n+2k+\nu+2}$, respectively, that were
not used for the construction of the Pad\'e approximants $\varepsilon_{2k}^{(n)}
= P_{k}^{k+n}$ or the transformation terms $\varphi_{2k}^{(n)}$.

In Table \ref{Table_3} we compute the rational function
$\varphi_{66}^{(0)} (\lambda^2)$ corresponding to the diagonal Pad\'e
approximant $\varepsilon_{66}^{(n)} = P_{33}^{33} (\lambda^2)$ with the
help of the recursive scheme (\ref{rec_phi}) from the coefficients
$b_{\nu}$ with $1 \le \nu \le 67$. The resulting expansion coefficients
$b_{\nu}^{(0, 66)}$ defined in (\ref{PhiSerPT}) with $\nu \ge 0$ provide
predictions to the coefficients $b_{\nu+68}$ of the perturbation series
(\ref{PerSerDelE0PT}). All calculations for Table \ref{Table_3} were
done free of rounding errors using the exact rational arithmetics of
Maple. Only in the final step the coefficients were converted to
floating point numbers for the sake of readability.

\begin{table}[htb]
\begin{center}
\parbox{5.30in}{\caption{\label{Table_3} Predictions $b_{n}^{(0, 66)}$ 
for the coefficients $b_{n+68}$ of the perturbation series
(\protect\ref{PerSerDelE0PT}) for $\Delta E_0 (\lambda^2)$ with $0 \le n
\le 15$.}} 
\begin{tabular*}{5.30in}{l@{\extracolsep{\fill}}rr} \\
[1\jot] \hline \hline \rule{0pt}{4\jot}%
$n$ & \multicolumn{1}{c}{$b_{n}^{(0, 66)}$} 
& \multicolumn{1}{c}{$b_{n+68}$} \\ 
[1\jot] \hline \rule{0pt}{4\jot}%
0  & $-0.118~625~502~281~564~111~353 \cdot 10^{217}$ & 
$-0.118~625~502~281~564~111~358 \cdot 10^{217}$ \\
1  & $ 0.487~707~952~691~623~584~397 \cdot 10^{220}$ & 
$ 0.487~707~952~691~623~585~158 \cdot 10^{220}$ \\
2  & $-0.203~437~822~070~101~216~978 \cdot 10^{224}$ & 
$-0.203~437~822~070~101~222~504 \cdot 10^{224}$ \\
3  & $ 0.860~803~267~021~875~481~138 \cdot 10^{227}$ & 
$ 0.860~803~267~021~875~756~369 \cdot 10^{227}$ \\
4  & $-0.369~393~498~548~727~222~559 \cdot 10^{231}$ & 
$-0.369~393~498~548~728~279~960 \cdot 10^{231}$ \\
5  & $ 0.160~732~212~082~002~560~522 \cdot 10^{235}$ & 
$ 0.160~732~212~082~005~901~209 \cdot 10^{235}$ \\
6  & $-0.709~026~471~212~486~114~145 \cdot 10^{238}$ & 
$-0.709~026~471~212~576~489~701 \cdot 10^{238}$ \\
7  & $ 0.317~020~667~799~578~470~271 \cdot 10^{242}$ & 
$ 0.317~020~667~799~793~728~631 \cdot 10^{242}$ \\
8  & $-0.143~648~198~373~426~854~924 \cdot 10^{246}$ & 
$-0.143~648~198~373~887~496~043 \cdot 10^{246}$ \\
9  & $ 0.659~514~281~085~804~565~498 \cdot 10^{249}$ & 
$ 0.659~514~281~094~798~452~336 \cdot 10^{249}$ \\
10 & $-0.306~750~687~264~795~900~309 \cdot 10^{253}$ & 
$-0.306~750~687~281~012~588~650 \cdot 10^{253}$ \\
11 & $ 0.144~514~693~689~642~646~364 \cdot 10^{257}$ & 
$ 0.144~514~693~716~909~093~737 \cdot 10^{257}$ \\
12 & $-0.689~498~329~409~437~387~151 \cdot 10^{260}$ & 
$-0.689~498~329~840~371~816~155 \cdot 10^{260}$ \\
13 & $ 0.333~104~548~293~054~144~923 \cdot 10^{264}$ & 
$ 0.333~104~548~937~521~023~558 \cdot 10^{264}$ \\
14 & $-0.162~924~769~352~053~020~131 \cdot 10^{268}$ & 
$-0.162~924~770~269~205~895~837 \cdot 10^{268}$ \\
15 & $ 0.806~654~532~549~091~198~441 \cdot 10^{271}$ & 
$ 0.806~654~545~029~445~531~410 \cdot 10^{271}$ \\
[1\jot] \hline \hline \rule{0pt}{4\jot}%
\end{tabular*}
\end{center}
\end{table}

The results in Table \ref{Table_3} show that the expansion coefficients
$b_{\nu}^{(0, 66)}$ with $\nu \ge 0$ provide already remarkably accurate
predictions to the corresponding coefficients $b_{\nu+68}$. Moreover,
the Stieltjes inequality (\ref{PadeMom_le_StieMom}) is satisfied in all cases.

In Table \ref{Table_4} we do the same calculations as in Table
\ref{Table_3}, but this time for the ground-state energy shift $\Delta
\mathcal{E}_0 (\beta)$ of the quartic anharmonic oscillator. Thus, the
expansion of the rational function $\varphi_{66}^{(0)} (\beta)$, which
is computed from the coefficients $\mathcal{B}_{\nu}$ with $1 \le \nu
\le 67$ of the perturbation series (\ref{PerSerDelE0QAHO}), provides 
predictions $\mathcal{B}_{\nu}^{(0, 66)}$ to the coefficients
$\mathcal{B}_{\nu+68}$.

\begin{table}[htb]
\begin{center}
\parbox{5.30in}{\caption{\label{Table_4} Predictions 
$\mathcal{B}_{n}^{(0, 66)}$ for the coefficients $\mathcal{B}_{n+66}$ of
the perturbation series (\protect\ref{PerSerDelE0QAHO}) for $\Delta E_0
({\beta})$ with $0 \le n \le 15$.}}
\begin{tabular*}{5.30in}{l@{\extracolsep{\fill}}rr} \\
[1\jot] \hline \hline \rule{0pt}{4\jot}%
$n$ & \multicolumn{1}{c}{$\mathcal{B}_{n}^{(0, 66)}$} 
& \multicolumn{1}{c}{$\mathcal{B}_{n+68}$} \\ 
[1\jot] \hline \rule{0pt}{4\jot}%
0  & $-0.243~941~384~991~118~295~771 \cdot 10^{108}$ & 
$-0.243~941~384~991~118~295~782 \cdot 10^{108}$ \\
1  & $ 0.250~725~042~695~070~353~544 \cdot 10^{110}$ & 
$ 0.250~725~042~695~070~353~955 \cdot 10^{110}$ \\
2  & $-0.261~457~030~278~874~510~535 \cdot 10^{112}$ & 
$-0.261~457~030~278~874~517~978 \cdot 10^{112}$ \\
3  & $ 0.276~569~040~522~183~341~803 \cdot 10^{114}$ & 
$ 0.276~569~040~522~183~434~367 \cdot 10^{114}$ \\
4  & $-0.296~701~814~375~736~021~569 \cdot 10^{116}$ & 
$-0.296~701~814~375~736~909~442 \cdot 10^{116}$ \\
5  & $ 0.322~749~390~515~363~534~568 \cdot 10^{118}$ & 
$ 0.322~749~390~515~370~538~244 \cdot 10^{118}$ \\
6  & $-0.355~923~577~678~312~610~650 \cdot 10^{120}$ & 
$-0.355~923~577~678~359~918~630 \cdot 10^{120}$ \\
7  & $ 0.397~845~013~388~761~856~087 \cdot 10^{122}$ & 
$ 0.397~845~013~389~043~208~304 \cdot 10^{122}$ \\
8  & $-0.450~670~140~529~734~425~361 \cdot 10^{124}$ & 
$-0.450~670~140~531~237~820~183 \cdot 10^{124}$ \\
9  & $ 0.517~267~603~130~982~724~472 \cdot 10^{126}$ & 
$ 0.517~267~603~138~312~496~560 \cdot 10^{126}$ \\
10 & $-0.601~463~530~952~366~420~452 \cdot 10^{128}$ & 
$-0.601~463~530~985~369~171~801 \cdot 10^{128}$ \\
11 & $ 0.708~383~831~448~709~420~419 \cdot 10^{130}$ & 
$ 0.708~383~831~587~280~706~922 \cdot 10^{130}$ \\
12 & $-0.844~934~259~800~460~726~512 \cdot 10^{132}$ & 
$-0.844~934~260~347~380~950~714 \cdot 10^{132}$ \\
13 & $ 0.102~047~769~191~033~928~036 \cdot 10^{135}$ & 
$ 0.102~047~769~395~298~233~145 \cdot 10^{135}$ \\
14 & $-0.124~779~572~617~585~209~351 \cdot 10^{137}$ & 
$-0.124~779~573~343~562~971~244 \cdot 10^{137}$ \\
15 & $ 0.154~446~315~576~951~606~777 \cdot 10^{139}$ & 
$ 0.154~446~318~044~174~985~350 \cdot 10^{139}$ \\ 
[1\jot] \hline \hline \rule{0pt}{4\jot}%
\end{tabular*}
\end{center}
\end{table}

A comparison of Tables \ref{Table_3} and \ref{Table_4} shows that their
qualitative features are identical. In this context it is quite
remarkable that although the coefficients $b_n$ grow significantly more
rapidly in magnitude than the coefficients $\mathcal{B}_n$, which
follows from the large-order estimates (\ref{e3}) and
(\ref{asy_cf_QAHO}), Pad\'e prediction nevertheless yields results of
virtually identical quality.

The Pad\'e prediction of unknown series coefficients based on the
recursive scheme (\ref{rec_phi}) is certainly computationally simpler
than the straightforward symbolic computation and expansion of Pad\'e
approximants. The rational function $\varphi_{2k}^{(n)} (z)$ has a
simpler structure than $\varepsilon_{2k}^{(n)} = P_{k}^{k+n} (z)$, and
the first $n+2k$ symbolic differentiations can be avoided. Nevertheless,
the recursive symbolic computation of the rational function
$\varphi_{2k}^{(n)} (z)$ from the coefficients $\gamma_{0}$,
$\gamma_{1}$, \ldots, $\gamma_{n+2k}$ of the power series
(\ref{PowSer_f}) can become quite demanding, in particular if $k$
becomes large.

The problems connected with the computation of $\varphi_{2k}^{(n)} (z)$
can largely be avoided if one only tries to compute a prediction
\begin{equation}
g_{2k}^{(n)} \; = \; g_{0}^{(n, 2k)}
% \label{}
\end{equation}
for the \emph{first} term $\gamma_{n+2k+1}$ not used for the construction
of either $\varepsilon_{2k}^{(n)}$ or $\varphi_{2k}^{(n)} (z)$. For that
purpose, we have only to set $z = 0$ in (\ref{rec_phi}). This yields the
following recursive scheme (Eq.\ (3.17) of \cite{We2000}):
\begin{subequations}
\label{Rec_g}
\begin{eqnarray}
g_{0}^{(n)} & = & 0 \, , \qquad n \in \mathbb{N}_0 \, ,
% \label{}
\\
g_{2}^{(n)} & = & \frac{\bigl[ \gamma_{n+2} \bigr]^2}{\gamma_{n+1}} \, ,
\qquad n \in \mathbb{N}_0 \, ,
% \label{}
\\
g_{2k+2}^{(n)} & = & g_{2k}^{(n+2)} \, + \, 
\frac
{\bigl[ \gamma_{n+2k+2} - g_{2k}^{(n+1)} \bigr]^2}
{\gamma_{n+2k+1} - g_{2k}^{(n)}} \, - \, 
\frac
{\bigl[ \gamma_{n+2k+2} - g_{2k}^{(n+1)} \bigr]^2}
{\gamma_{n+2k+1} - g_{2k-2}^{(n+2)}} \, , \nonumber \\
& & k \in \mathbb{N} \, , \qquad n \in \mathbb{N}_0 \, .
% \label{}
\end{eqnarray}
\end{subequations}
The main advantage of this recursive scheme over the recursive scheme
(\ref{rec_phi}), from which it was derived, is that it only involves
numbers but no symbolic expressions. 

\renewcommand{\topfraction}{0.8}

\begin{table}[t]
\begin{center}
\parbox{4.00in}{\caption{\label{Table_5} Relative errors $R_n$ and
$\mathcal{R}_n$ defined in (\protect\ref{RelErr_b}) and
(\protect\ref{RelErr_B}) of the Pad\'e predictions for the first
coefficients of the perturbation series (\protect\ref{PerSerDelE0PT})
and (\protect\ref{PerSerDelE0QAHO}) not used for the construction of the
rational approximants.}}
\begin{tabular*}{4.00in}{l@{\extracolsep{\fill}}ll} \\
[1\jot] \hline \hline \rule{0pt}{4\jot}%
$n$ 
& \multicolumn{1}{c}{$R_n$} 
& \multicolumn{1}{c}{$\mathcal{R}_n$} \\
[1\jot] \hline \rule{0pt}{4\jot}%
2   & $-0.295~410~699$ & $-0.316~117~394$ \\                   
3   & $-0.207~610~910$ & $-0.218~823~682$ \\               
4   & $-0.759~683~860 \cdot 10^{-1}$  
& $-0.833~341~229 \cdot 10^{-1}$ \\ 
5   & $-0.483~909~816 \cdot 10^{-1}$  
& $-0.522~231~970 \cdot 10^{-1}$ \\ 
6   & $-0.197~254~000 \cdot 10^{-1}$  
& $-0.218~362~310 \cdot 10^{-1}$ \\ 
7   & $-0.120~176~754 \cdot 10^{-1}$  
& $-0.130~908~858 \cdot 10^{-1}$\\ 
[1\jot] \hline \rule{0pt}{4\jot}%
50  & $-0.258~379~657 \cdot 10^{-14}$ 
& $-0.273~374~025 \cdot 10^{-14}$ \\
51  & $-0.134~007~443 \cdot 10^{-14}$ 
& $-0.141~313~446 \cdot 10^{-14}$ \\
52  & $-0.658~949~507 \cdot 10^{-15}$ 
& $-0.696~514~376 \cdot 10^{-15}$ \\
53  & $-0.341~294~856 \cdot 10^{-15}$ 
& $-0.359~600~798 \cdot 10^{-15}$ \\
54  & $-0.167~932~626 \cdot 10^{-15}$ 
& $-0.177~342~673 \cdot 10^{-15}$ \\
[1\jot] \hline \rule{0pt}{4\jot}%
100 & $-0.327~674~717 \cdot 10^{-29}$ 
& $-0.341~449~758 \cdot 10^{-29}$ \\
101 & $-0.166~900~610 \cdot 10^{-29}$ 
& $-0.173~636~626 \cdot 10^{-29}$ \\
102 & $-0.827~630~688 \cdot 10^{-30}$ 
& $-0.862~097~827 \cdot 10^{-30}$ \\
103 & $-0.421~400~183 \cdot 10^{-30}$ 
& $-0.438~256~801 \cdot 10^{-30}$ \\
104 & $-0.208~999~229 \cdot 10^{-30}$ 
& $-0.217~623~211 \cdot 10^{-30}$ \\
[1\jot] \hline \rule{0pt}{4\jot}%
140 & $-0.354~821~178 \cdot 10^{-41}$ 
& $-0.367~523~224 \cdot 10^{-41}$ \\
141 & $-0.179~775~658 \cdot 10^{-41}$ 
& $-0.186~000~741 \cdot 10^{-41}$ \\
142 & $-0.893~606~384 \cdot 10^{-42}$ 
& $-0.925~378~776 \cdot 10^{-42}$ \\
143 & $-0.452~675~070 \cdot 10^{-42}$ 
& $-0.468~247~723 \cdot 10^{-42}$ \\
144 & $-0.225~028~932 \cdot 10^{-42}$ 
& $-0.232~976~269 \cdot 10^{-42}$ \\
[1\jot] \hline \rule{0pt}{4\jot}%
187 & $-0.295~089~701 \cdot 10^{-55}$ 
& $-0.304~045~084 \cdot 10^{-55}$ \\
188 & $-0.146~895~157 \cdot 10^{-55}$ 
& $-0.151~456~461 \cdot 10^{-55}$ \\
189 & $-0.741~750~230 \cdot 10^{-56}$ 
& $-0.764~148~505 \cdot 10^{-56}$ \\
190 & $-0.369~259~543 \cdot 10^{-56}$ 
& $-0.380~666~936 \cdot 10^{-56}$ \\
191 & $-0.186~438~979 \cdot 10^{-56}$ 
& $-0.192~040~968 \cdot 10^{-56}$ \\
[1\jot] \hline \hline \rule{0pt}{4\jot}%
\end{tabular*}
\end{center}
\end{table}

In Table \ref{Table_5} we present selected results for the Pad\'e
predictions of the \emph{first} coefficients of the perturbation
expansion (\ref{PerSerDelE0PT}) for $\Delta E_0 (\lambda^2)$ and
(\ref{PerSerDelE0QAHO}) for $\Delta \mathcal{E}_{0} (\beta)$, which were
not used in the Pad\'e approximants $\varepsilon_{2 \Ent{n/2}}^{(n - 2
\Ent{n/2})} = P_{\Ent{n/2}}^{n-\Ent{n/2}}$ for $2 \le n \le 191$. The
first predictions $b_{0}^{(n-2\Ent{n/2}, 2\Ent{n/2})}$ to $b_{n+2}$ and
$\mathcal{B}_{0}^{(n-2\Ent{n/2}, 2\Ent{n/2})}$ to $\mathcal{B}_{n+2}$
were computed with the help of the recursive scheme (\ref{Rec_g}). For
the sake of readability, we present in Table \ref{Table_5} only the
relative errors
\begin{equation}
R_n \; = \; 
\frac{b_{0}^{(n-2\Ent{n/2}, 2\Ent{n/2})} - b_{n+2}}{b_{n+2}} 
\label{RelErr_b}
\end{equation}
and 
\begin{equation}
\mathcal{R}_n \; = \;
\frac{\mathcal{B}_{0}^{(n-2\Ent{n/2}, 2\Ent{n/2})} - 
\mathcal{B}_{n+2}}{\mathcal{B}_{n+2}} \, .
\label{RelErr_B}
\end{equation}
If the input data of the recursive scheme (\ref{Rec_g}) are exact
rational numbers as the coefficients $b_n$ or $\mathcal{B}_n$, then the
predictions can be computed free of rounding errors. However, it turned
out that that the predictions computed in this way were \emph{huge}
rational numbers which slowed down computation considerably. Therefore,
we used the floating-point arithmetics of Maple with an accuracy of 600
decimal digits for the computation of results presented in Table
\ref{Table_5}.

The results in Table \ref{Table_5} show that the first coefficients not
used for the construction of the Pad\'e approximants $\varepsilon_{2
\Ent{n/2}}^{(n - 2 \Ent{n/2})} = P_{\Ent{n/2}}^{n-\Ent{n/2}}$ can be
predicted with remarkable accuracy by the recursive scheme (\ref{Rec_g})
if $n$ is sufficiently large. Moreover, the agreement of the relative
errors $R_n$ and $\mathcal{R}_n$ is stunning. However, for our purposes 
most important is the observation that all relative errors in Table
\ref{Table_5} are negative which is in agreement with the Stieltjes
inequality (\ref{PadeMom_le_StieMom}). 

Thus, the prediction results of this section show that there is no
significant difference between the prediction of coefficients $b_n$ of
the perturbation series (\ref{PerSerDelE0PT}), whose Stieltjes nature we
want to establish, and the prediction of the coefficients
$\mathcal{B}_n$ of the analogous quartic anharmonic oscillator
perturbation series (\ref{PerSerDelE0QAHO}), whose Stieltjes nature was
established rigorously by Simon (Theorem IV.2.1 of
\cite{Si1970}). Accordingly, these results provide further numerical
evidence that the perturbation series (\ref{PerSerDelE0PT}) for the
ground-state energy shift $\Delta E_0 (\lambda^2)$ is indeed a Stieltjes
series.

\typeout{==> Section 5} 
\section{Summary and Conclusions} 
\label{Sec:SumConclu}

As shown by countless monographs or articles, the mathematical theory of
the conventional Hermitian Hamiltonians of quantum mechanics is well
established and has reached a high degree of sophistication. Moreover,
the divergence and the summation of the perturbation expansions
resulting from these Hamiltonians is also comparatively well understood,
in particular in the case of the Pad\'e summation of Stieltjes
series. In contrast, the rigorous mathematical theory of non-Hermitian,
${\cal PT}$-symmetric Hamiltonians is virtually nonexistent. It is not
known whether divergent perturbation expansions resulting from such
Hamiltonians can be summed to yield a uniquely determined
result. Consequently, the best we can do for the moment is to perform
numerical studies from which we can try to draw general qualitative
conclusions.

The main intention of this paper is to provide numerical evidence that the
perturbation series (\ref{PerSerDelE0PT}) for the ground-state energy shift
$\Delta E_0(\lambda^2)$ is a Stieltjes series, because this would guarantee that
certain subsequences of the Pad\'e table constructed from the partial sums of
this perturbation series converge to a uniquely determined summation result, as
discussed in Section \ref{Sec:PadEpsStie}.

If the Pad\'{e} approximants are computed with the help of Wynn's recursive
algorithm (\ref{eps_al}) according to (\ref{Eps_Pade}) and (\ref{EpsPadeAppSeq})
-- as it is done in this paper -- and if the series to be transformed is the
moment expansion of a Stieltjes function, then the Pad\'{e} approximants must
satisfy inequality (\ref{EpsPadeIneq}). As shown in Table \ref{Table_2}, the
Pad\'{e} summation results for the perturbation series (\ref{PerSerDelE0PT}) for
the ground-state energy shift $\Delta E_0 (\lambda^2)$ as well as for the
analogous perturbation series (\ref{PerSerDelE0QAHO}) for the ground-state
energy shift $\Delta\mathcal{E}_{0} (\beta)$ of the quartic anharmonic
oscillator, which is known to be a Stieltjes series, satisfy this inequality.
Moreover, the divergence of the two perturbation expansions
(\ref{PerSerDelE0PT}) and (\ref{PerSerDelE0QAHO}) as well as as the convergence
of their Pad\'{e} summation results is virtually indistinguishable if the two
coupling constants $\lambda$ and $\beta$ satisfy (\ref{beta2lambda}).

If a Pad\'{e} approximant $P_{m}^{l} (z)$ constructed from the partial sums
(\ref{ParSumStieSer}) of the moment expansion for a Stieltjes function $F(z)$ is
expanded in power series around $z=0$ according to (\ref{PowSerStiePade}), then
this series must be strictly alternating for $z>0$ just like the Stieltjes
series for $F (z)$. Moreover, the coefficients $\mu_{n}^{[l/m]}$ of this
expansion are bounded in magnitude by the Stieltjes moments $\mu_n$ according to
the inequality (\ref{PadeMom_le_StieMom}).

Thus, via the inequality (\ref{PadeMom_le_StieMom}) it can be checked whether a
moment expansion of the type of (\ref{StieSer}) is a Stieltjes series. However,
the accuracy-through-order relationship (\ref{OestPA}) implies that this
inequality is by default satisfied for all indices $n\le l+m$, and only for
$n\ge l+m+1$ do we obtain useful information. In particular for large values of
$l$ and $m$, the symbolic construction of Pad\'{e} approximants and their
subsequent expansion may become quite demanding both with respect to time and
memory. 

These computational problems can to some extent be overcome by expressing the
Pad\'{e} approximants $P_{k}^{k+n}(z)$, that can be computed with the help of
Wynn's epsilon algorithm, according to (\ref{EpsRem_phi}) by the partial sum
from which it was constructed plus the transformation term $z^{n+2k+1}
\varphi_{2k}^{(n)} (z)$. The quantities $\varphi_{2k}^{(n)} (z)$ can be computed
recursively with the help of (\ref{rec_phi}), and their computation is less
demanding than the recursive computation of the corresponding Pad\'{e}
approximants $P_{k}^{k+n} (z)$. Moreover, the Taylor expansion of
$\varphi_{2k}^{(n)}(z)$ according to (\ref{PowSerPhi}) yields the desired
coefficients $\mu_{n+2k+\nu+1}^{[k+n/k]}$ with $\nu \ge 0$.

In Tables \ref{Table_3} and \ref{Table_4}, we proceed as described above and
confirm the validity of the Stieltjes inequality (\ref{PadeMom_le_StieMom}) in
the case of the first 15 coefficients of the quantities $\varphi_{66}^{(0)}
(\lambda^2)$ and $\varphi_{66}^{(0)}(\beta)$ corresponding to the Pad\'{e}
approximants $P_{33}^{33} (\lambda^2)$ and $P_{33}^{33} (\beta)$, respectively.
In this context, it is remarkable that the quality of the predictions is
virtually indistinguishable although the coefficients $b_n$ of the
$\mathcal{PT}$-symmetric perturbation series grow much more rapidly than
the coefficients $\mathcal{B}_n$ for the quartic anharmonic oscillator.

For larger values of $n$ and $k$ the symbolic computation of the quantities
$\varphi_{2k}^{(n)} (z)$ via the recursive scheme (\ref{rec_phi}) becomes quite
demanding. In such a case, it is much simpler to compute only the prediction for
the first series coefficient not used for the computation of $\varphi_{2k}^{(n)}
(z)$. This can be done with the help of the recursive scheme (\ref{Rec_g}). In
Table \ref{Table_5} we show that all first predictions, which we can compute
from the coefficients $b_{\nu}$ and $\mathcal{B}_{\nu}$ with $1\le \nu \le 192$,
satisfy the Stieltjes inequality (\ref{PadeMom_le_StieMom}). Moreover, the
quality of the prediction results is again virtually identical. 

We are of course aware that numerical results cannot replace rigorous
mathematical proofs. Nevertheless, we believe that our numerical experiments are
both interesting and useful, and that they provide considerable evidence that
the perturbation series (\protect\ref{PerSerDelE0PT}) for the ground-state
energy shift $\Delta E_0(\lambda^2)$ of the $\mathcal{PT}$-symmetric Hamiltonian
(\ref{e1}) is indeed a Stieltjes series, which would imply that the Pad\'{e}
summation of this divergent perturbation series converges.

\typeout{==> Acknowledgment}
\section* {Acknowledgment}

CMB is grateful to the U.S.~Department of Energy for financial support and EJW
thanks the Fonds der Chemischen Industrie for financial support. 

\typeout{==> References}

\end{document}